\documentclass[12pt]{article}
\usepackage{epsfig}
\newenvironment{drfigs}[1]{\par\vspace{#1}\ignorespaces}
{\par\vspace{12pt}\ignorespaces}
\title{\bf Optimal Signal-to-Quantum Noise Ratio for Nonclassical Number States}
\author{
Yinqi Feng\footnote{Email: y.feng@open.ac.uk}$\;\;$  and Allan I. Solomon\footnote{Email: a.i.solomon@open.ac.uk}\\
{ \small \sl Faculty of Mathematics and Computing, The Open 
University,
Milton Keynes, MK7 6AA, UK } \\
}
\begin{document}
\maketitle
\begin{abstract}
We analyze the properties of nonclassical number states, specifically squeezed number states $D(\alpha)S(z)\vert n \rangle$, and find their maximum signal-to-quantum noise ratio. It is shown that the  optimal signal-to-quantum noise ratio for these states decreases as $1/(2n+1)^2$, where $n$ is the photon number, from the optimal value as  derived by Yuen.  
\end{abstract}
\section{Introduction}
There has been a great deal of interest recently in nonclassical 
number states of the electromagnetic 
field.  In addition to the unusual physical properties they possess 
\cite{kim}-\cite{nieto1} there is the hope that in the not too distant 
future these number states can be observed \cite{matos}-\cite{nieto2}
and, if experimentally generated,  lead to significant 
improvement in optical system performance. 
In his classic paper of 1976, Yuen \cite{yuen}  proved the following important result:
\begin{itemize}
\item   for an 
arbitrary quantum state of radiation with frequency  
$\omega$, the optimum signal-to-quantum noise ratio $\rho$  
for fixed energy (or power per unit 
frequency ) $\hbar\omega N_s$ has the value $4N_s(N_s+1)$.
\item he additionally showed that this optimal value was attainable by 
the squeezed coherent vacuum.
\end{itemize}
This result, which is both elementary - in that it depends only on the basic postulates of quantum mechanics, and general - in that it  applies both to states and density functions, has recently been extended to the case of deformed photons \cite{sol}. The question naturally arises as to what extent the optimal value must be modified for other conventional photon states.  In our paper we derive  the optimum signal-to-quantum noise ratios for nonclassical number states, in particular  squeezed number states. These may be taken as a paradigm for the squeezing of an imperfect vacuum; that is, when there are a small number of photons present, we find for   photon  number $n$, the ratio $\rho$  for squeezed number states decreases sharply - as $1/(2n+1)^2$ -  from  Yuen's result $ 4N_s(N_s+1) $, which  is  only attainable for the usual  coherent squeezed state.
\section{Coherent Number states and Squeezed Number states}
In this section we summarize some properties of coherent number 
states and squeezed 
number states necessary in caculating  the signal-to-quantum 
noise ratio.
Coherent number states are defined by 
\begin{equation}
| \alpha, n > = D(\alpha) | n >
\label{a} 
\end{equation}
where $D(\alpha)$ is the displacement operator \cite{man},  given by
\begin{equation}
D(\alpha) = \exp (\alpha a^{\dagger} - \alpha^{*} a )
\label{b}
\end{equation}
For $n=0$, the coherent number state $ |\alpha, n>$ reduces to the 
usual coherent state $D(\alpha) |0>$.
The displacement operator produces the  
transformations \cite{man}:
\begin{equation}
D^{\dagger}(\alpha) a D(\alpha) = a +\alpha  \; \; \; \; 
D^{\dagger}(\alpha) a^{\dagger}  D(\alpha)= a^{\dagger} + 
\alpha^{*}
\label{c}
\end{equation}
Quadrature  operators are defined as usual  by 
\begin{equation}
X =  \frac{1}{\sqrt{2}}  ( a + a^{\dagger}) \; \; \; \; 
P =  \frac{1}{i \sqrt{2}}( a - a^{\dagger}). 
\label{d}
\end{equation}
With the use of Equations (\ref{a}) -- (\ref{d}) , we easily obtain the 
following mean values in the 
state $|\alpha, n >$:
\begin{eqnarray}
\langle N \rangle &=&n +  \vert \alpha \vert ^2 \nonumber  \; \; \; \; (N\equiv a^{\dagger}a )\\
\langle X \rangle &=&  \frac{1}{\sqrt{2}}  (\alpha   + \alpha^{*})  
\nonumber\\
\langle P \rangle &=&  \frac{1}{i \sqrt{2}}  (\alpha -  \alpha^{*}) \label{e}   
\\
 (\Delta X )^2 _ n &=& \frac{1}{2} ( 2 n   +  1 )  \nonumber\\
 (\Delta P )^2 _n &=& \frac{1}{2} ( 2 n   +  1 )    \nonumber
\end{eqnarray}
where $  (\Delta X )^2  =\langle  X ^2 \rangle - 
\langle  X   \rangle ^ 2 , \;\; \;  (\Delta P )^2 = 
\langle  P^ 2 \rangle - \langle P  \rangle ^ 2. $\\
Squeezed number states \cite{nieto1} are defined by
\begin{equation}
| z, \alpha, n > = D(\alpha) S ( z )  | n > 
\label{f}
\end{equation}
where $S ( z )$ is the squeezing operator, given by
\begin{equation}
S ( z ) =\exp[\frac{1}{2}( z a^{\dagger2} - z^{*} a^2 )] ,  \; \; \; 
 z = r\exp ( i\phi )
\label{g}
\end{equation}
Note that this 2-parameter operator is not the most general squeezing operator, which is a 3-parameter element of the group SU(1,1). $D(\alpha)$ is the displacement operator of Eq. (\ref{b}). For $ n= 0$  the squeezed number states
$| z, \alpha ,n> $
reduce to the more familiar  squeezed  states $ D(\alpha) S(z) | 0 >$. The unitary transformation of the operators $a$ and $a^\dagger$ by  $S(z)$ 
and $S^{\dagger}(z)$ is given by \cite{man}:
\begin{eqnarray}
S^{\dagger}(z) a S(z)   &=&\lambda a + \mu a^{\dagger}   \nonumber \\
S^{\dagger}( z ) a^{\dagger} S( z ) &=&\lambda a^{\dagger}
+ \mu ^{*} a
\label{h}
\end{eqnarray}
where we have put 
\[
\lambda= \cosh r, \; \; \;  \mu= \exp(i \phi ) \sinh r  
 \]
So, with the use of Equations (\ref{c}) -(\ref{d})  and (\ref{f})
-(\ref{h})  we can obtain the following mean 
values in the state $ | z, \alpha, n >$:
\begin{eqnarray}
\langle N \rangle & =&   |\lambda|^2 n + |\mu|^2 (n + 1) +|\alpha |^2 \nonumber \\
\langle X \rangle & =& \frac{1}{\sqrt{2}}(\alpha + \alpha^{*})  \nonumber\\
\langle P \rangle& =& \frac{1}{i \sqrt{2}}(\alpha - \alpha^{*})\label{i}\\
 (\Delta X )^2 _n & =& \frac{1}{2} |\lambda + \mu |^2 ( 2 n   +  1 
)  \nonumber\\
 (\Delta P )^2 _n &=&  \frac{1}{2}  |\lambda- \mu |^2 ( 2 n   +  1 )\nonumber  
\end{eqnarray}
\section{The Signal-to-Quantum Noise Ratio $\rho$ }
The signal-to-quantum noise ratio  $\rho_{|> }$ in the state $ |> $ is defined by
\begin{equation}
\rho _{|>} = \frac{\langle X \rangle^2}{ (\Delta X )^2 }
\label{j}
\end{equation}
So for the coherent number state $ |\alpha, n >$, we have
\begin{equation}
\rho_{ |\alpha, n >}=\frac{4(Re\alpha)^2}{2n+1}
\label{k}
\end{equation}
and for the squeezed number state $| z, \alpha, n >$, we have
\begin{equation}
\rho _{ | z, \alpha, n > }=\frac{|\alpha+ \alpha^{*}|^2}{|\lambda + \mu|^2 (2n+1)}.
\label{l}
\end{equation}
Note that $X^2 + P^2 = 2N+1$, so that  $\langle X^2 \rangle  + \langle P^2 \rangle  =\langle  2N+1\rangle$, whence 
\begin{equation} 
\langle X \rangle ^2 +(\Delta X)^2 + \langle P \rangle ^2 +(\Delta P)^2 =\langle  2N+1\rangle .
\end{equation}
Under the energy (or power per unit frequency) constraint \cite{yuen}
$$  \hbar \omega \langle N \rangle \leq   \hbar \omega  N_s $$
the signal-to-quantum noise ratio can be maximised by using all the available energy 
and allocating  no energy to  $\langle P \rangle$;
 that is 
\begin{equation}
\langle N \rangle =  N_s,\;\;\;  \langle P \rangle = 0 .
\end{equation}
The expression Eq. (\ref{j}) becomes
\begin{equation}
\rho_{|>} = \frac{(2 N_s +1) - (\Delta X)^2 - (\Delta P)^2}{(\Delta X)^2}.
\label{rho3}
\end{equation}
Using  the uncertainty relation
\begin{equation}
 (\Delta X )^2   (\Delta P )^2 
 \geq \frac{1}{4}(2n+1)^2
\end{equation}
 we can optimize Eq. (\ref{rho3})   in terms of $ (\Delta X )^2 $ alone:
\begin{equation}
\rho_{ | >} = \frac{(2 N_s +1)}{ (\Delta X)^2} -1 - \frac{(n+\frac{1}{2})^2}{(\Delta X)^2}.
\end{equation}
We thus find that the maximum value is given by 
\begin{equation}
\rho_{ | \alpha, n >max} = \frac{4(N_s-n)}{2n+1}  \label{p}
\end{equation}
for  the coherent number state $ | \alpha, n >$  with $\alpha 
=\sqrt{N_s-n}$  , and
\begin{equation}
 \rho_{ | z, \alpha, n > max } = \frac{4( N_s-n) (N_s+1+n)}{(2n+1)^2}
\label{q}
\end{equation}
obtained with the squeezed number state $ | z, \alpha, n >$ with 
\begin{eqnarray}  
\alpha &=& \sqrt{\frac{(N_s-n)(N_s+1+n)}{(2N_s+1)}} \nonumber\\
\lambda & =& \frac{N_s+1+n}{\sqrt{(2N_s+1)(2n+1)}} \\
\mu & =&  \frac{n-N_s}{\sqrt{(2N_s+1)(2n+1)}}\nonumber
\end{eqnarray}
For the case where $n=0$ in Eqs. (\ref{p}) and (\ref{q}) , one finds the same 
result as Yuen \cite{yuen}.
\section{ Discussion}
From Eqs.(\ref{e}) and (\ref{i}), we obtain
\begin{equation}
 (\Delta X )^2 _n  (\Delta P )^2 
_n=\frac{1}{4} (2n+1)^2
\end{equation}
for coherent number states and 
\begin{equation}
 (\Delta X )^2 _n  (\Delta P )^2 
_n=\frac{1}{4}|\lambda^2-\mu^2|^2 (2n+1)^2
\end{equation}
for squeezed number states.  So we obtain in both cases
$$  (\Delta X )^2 _n  (\Delta P )^2 _n 
\geq  (\Delta X )^2 _0  (\Delta P )^2 
_0=\frac{1}{4}$$
This means that the number states are not ordinary 
minimum uncertainty states 
except for $n = 0$.
For Eqs. (\ref{k}) and (\ref{l}), we also have
\begin{eqnarray}
\rho_{ | \alpha, n >} <   \rho  _{| \alpha >}\nonumber\\
\rho _{ | z , \alpha, n >} <   \rho_{| z , \alpha >} \nonumber
\end{eqnarray}
for given complex values of  $\alpha$ and $z$. From Equation (\ref{q}) we obtain
\begin{equation}
\frac{\rho _{ | z , \alpha, n >max}}{\rho _{ | z , \alpha >max}} = \frac{(N_s-n)(N_s+1+n)}{N_s(N_s+1)(2n+1)^2} = \frac{1}{(2n+1)^2}[1-\frac{n(n+1)}{N_s(N_s+1)}]
\end{equation}
This ratio is plotted in Fig.1. We can see that in terms of the photon number $n$, the ratio decreases 
as $\frac{1}{(2n+1)^2}$ (for any $N_s>>n$), whence a slight deviation from a squeezed coherent vacuum results in large diminution of the optimal signal-to-quantum noise ratio $\rho_n$.\\
We note that for the states $S( z ) | n >$ (c.f. \cite{kim} and \cite{oliv}), which are obtained by letting $\alpha = 0$ in Eqs. (\ref{f}) and (\ref{i}), the optimal ratio is $ \rho_{ S(z)| n>}  = 0. $ Therefore as far as the signal-to-noise properties are concerned, the states of the form $S( z ) | n > $ may lead to bad optical system performance.

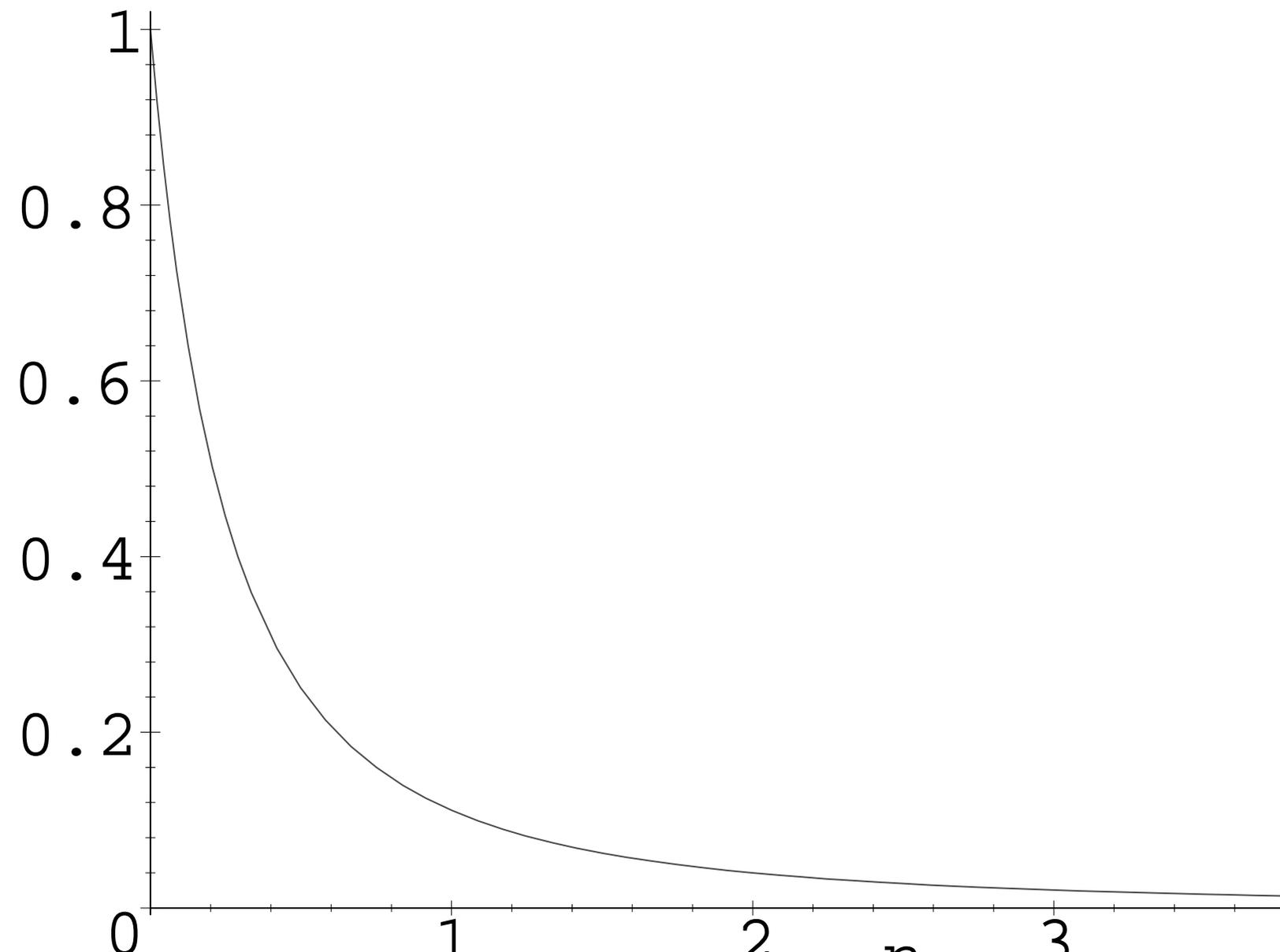
\end{document}